\documentclass[twocolumn,showpacs,preprintnumbers,amsmath,amssymb,nofootinbib]{revtex4}
\input psfig.sty
\def \be {\begin{equation}}
\def \ee {\end{equation}}
\def \bea {\begin{eqnarray}}
\def \eea {\end{eqnarray}}

\usepackage{graphicx}
\usepackage{dcolumn}
\usepackage{bm}

\begin{document}

\title{Scalar-field-dominated cosmology with a transient accelerating phase}

\author{F. C. Carvalho$^{1}$}
\email{fabiocc@on.br}

\author{J. S. Alcaniz$^{1}$}
\email{alcaniz@on.br}

\author{J. A. S. Lima$^{2}$}
\email{limajas@astro.iag.usp.br}

\author{R. Silva$^{1,3}$}
\email{rsilva@on.br}

\affiliation{$^{1}$Observat\'orio Nacional, Rua Gal. Jos\'e
Cristino 77, 20921-400 Rio de Janeiro, RJ, Brasil}

\affiliation{$^{2}$IAG, Universidade de S\~ao Paulo, 05508-900
S\~ao Paulo, SP, Brasil}

\affiliation{$^{3}$Universidade do Estado do Rio Grande do Norte,
59610-210, Mossor\'o, RN, Brasil}

\date{\today}


\begin{abstract}
A new cosmological scenario driven by a slow rolling homogeneous
scalar field whose exponential potential $V(\Phi)$ has a quadratic
dependence on the field $\Phi$ in addition to the standard linear
term is discussed. The derived equation of state for the field
predicts a transient accelerating phase, in which the Universe was
decelerated in the past, began to accelerate at redshift $z \sim 1$, is currently accelerated, but, finally, will return
to a decelerating phase in the future. This overall dynamic
behavior is profoundly different from the standard $\Lambda$CDM
evolution, and may alliviate some conflicts in reconciling the
idea of a dark energy-dominated universe with observables in
String/M-theory. The theoretical predictions for the present
transient scalar field plus dark matter dominated stage are
confronted with cosmological observations in order to test the
viability of the scenario.
\end{abstract}

\maketitle

\emph{Introduction.}---The idea of a dark energy-dominated
universe is a direct consequence of a convergence of independent
observational results, and constitutes one of the greatest
challenges for our current understanding of fundamental physics
\cite{review}. Among a number of possibilities to describe this
dark energy component, the simplest and most theoretically
appealing way is by means of a cosmological constant $\Lambda$,
which acts on the Einstein field equations as an isotropic and
homogeneous source with a constant equation of state (EoS) $w
\equiv p/\rho = -1$. Although cosmological scenarios with a
$\Lambda$ term may explain most of the current astronomical
observations, from the theoretical viewpoint it is really
difficult to reconcile the small value required by observations
($\simeq 10^{-10} \rm{erg/cm^{3}}$) with estimates from quantum
field theories ranging from 50-120 orders of magnitude larger
\cite{weinberg}, which makes a complete cancellation (from some
unknown string theory symmetry) seem also plausible.

However, if the cosmological term is null or it is not decaying in
the course of the expansion \cite{decaying},  something else must
be causing the Universe to speed up. The next simplest approach
toward constructing a model for an accelerating universe is to
work with the idea that the unknown, unclumped dark energy
component is due exclusively to a minimally coupled scalar field
$\Phi$ (quintessence field) which has not yet reached its ground
state and whose current dynamics is basically determined by its
potential energy $V(\Phi)$. This idea has received much attention
over the past few years and a considerable effort  has been made
in understanding the role of quintessence fields on the dynamics
of the Universe \cite{quintessence}. Examples of quintessence
potentials are ordinary exponential functions $V(\Phi) =
V_0\exp{(-\lambda \Phi)}$ \cite{RatraPeebles,fj,wetterich},
simple power-laws of the type $V(\Phi) = V_0\Phi^{-n}$
\cite{power_law},  combinations of exponential and sine-type
functions $V(\Phi) = V_0\exp{(-\lambda \Phi)}[1 + A\sin{(-\nu
\Phi)}]$ \cite{dodelson}, among others (see, e.g., \cite{review}
and references therein). In particular, the exponential example
above, originally investigated in Ref. \cite{RatraPeebles},
constitutes a kind of benchmark of quintessence scenarios and has
been largely explored in the literature, both in its theoretical
and observational aspects \cite{fj}. As shown in Ref.
\cite{wetterich} this particular class of potentials also leads to
an attractor-type solution with $\Omega_{\Phi} =
\rho_{\Phi}/(\rho_{\Phi} + \rho_i) = n/\lambda^2$, where
$\Omega_{\Phi}$ and $\rho_i$ are, respectively, the scalar field
density parameter and the energy density of the other component
scaling as $a^{-n}$. All these quintessence scenarios are based on
the premise that fundamental physics provides motivation for light
scalar fields in nature so that a quintessence field $\Phi$ may
not only be identified as the dark component dominating the
current cosmic evolution but also as a bridge between an
underlying theory and the observable structure of the Universe.

If, however, it is desirable (and we believe so) a more complete
connection between the physical mechanism behind dark energy and a
fundamental theory of nature, one must bear in mind that an
eternally accelerating universe, a rather generic feature of
quintessence scenarios, seems not to be in agreement with
String/M-theory predictions, since it is endowed with a
cosmological event horizon which prevents the construction of a
conventional S-matrix describing particle interactions
\cite{fischler}. Although the transition from an initially
decelerated to a late-time accelerating expansion is becoming
observationally established \cite{ms}, the duration of the
accelerating phase, depends crucially on the cosmological scenario
and, several models, which includes our current standard
$\Lambda$CDM scenario, imply an eternal acceleration or even an
accelerating expansion until the onset of a cosmic singularity
(e.g., the so-called phantom cosmologies \cite{ph}). This dark
energy/String theory conflict, therefore, leaves us with the
formidable task of either finding alternatives to the conventional
S-matrix (or, equivalently, defining observables in a string
theory described by a finite dimensional Hilbert space) or
constructing a quintessence model of the Universe that predicts
the possibility of a transient acceleration phenomenon.

In this \emph{Letter},  we follow the latter route and investigate
a new quintessence scenario driven by a rolling homogeneous scalar
field whose exponential potential $V(\Phi)$ predicts a transient
accelerating phase followed by an eternally decelerated universe.
The potential, which has a quadratic dependence on the field
$\Phi$ in addition to the standard linear term, is obtained
through a simple \emph{ansatz} and fully reproduces the exponential potential studied by Ratra
and Peebles in Ref. \cite{RatraPeebles} in the limit of the
dimensionless parameter $\alpha \rightarrow 0$. For all values of
$\alpha \neq 0$, however, the potential is dominated by the
quadratic contribution present in the exponential function,
admiting a wider range of solutions. We also derive analytically
all the main background equations of the model to show that a
transient accelerating phase is a feature of this class of
potentials, which in turn may reconcile the observed acceleration
of the Universe with the requirements of String/M theories. The
observational viability of our model is also tested by confronting
its predictions with the most recent SNe Ia and Cosmic Microwave
Background (CMB) data.

\emph{The Model.}---Let us first consider a homogeneous,
isotropic, spatially flat cosmologies described by the
Friedmann-Robertson-Walker (FRW) flat line element,
$ds^2=-dt^2+a^2(t)(dx^2+dy^2+dz^2),$
where $a(t)$ is the cosmological scalar factor and we have set the
speed of light $c = 1$. The action for the model is given by 
$S={m^2_{pl}/16\pi}\int d^4 x \sqrt{-g}[R
-{1\over2}\partial^{\mu}\Phi\partial_{\mu}\Phi-V(\Phi)+{\cal{L}}_{m}]$,
where $R$ is the Ricci scalar and $m_{pl}\equiv G^{-1/2}$ is
the Planck mass. The scalar field is assumed to be homogeneous,
such that $\Phi=\Phi(t)$ and the Lagrangian density
${\cal{L}}_{m}$ includes all matter and radiation fields.

\vspace{0.1cm}

{\centerline{\emph{1. A scalar-field-dominated universe.}}}

\vspace{0.1cm}

For now, it will be assumed that the cosmological fluid is
composed only of a quintessence field $\Phi$ (${\cal{L}}_{m} =
0$), whose energy-momentum tensor reads
$T^{\mu\nu}_{\Phi}=\partial^{\mu}\Phi\partial^{\nu}\Phi
-{1\over2}g^{\mu\nu}\left[\partial^{\alpha}\Phi\partial_{\alpha}\Phi
+2V(\Phi)\right]\;$.
The conservation equation for this $\Phi$ component takes the form
\be
\label{eq_cons} \dot\rho_{\Phi}+3H(\rho_{\Phi}+p_{\Phi})=0\;, \ee
or, equivalently,
$\ddot\Phi+3H\dot\Phi+V'(\Phi)=0$,
where $\rho_{\Phi}={1\over2}\dot\Phi^2+V(\Phi)$ and
$p_{\Phi}={1\over2}\dot\Phi^2-V(\Phi)$ are, respectively, the
scalar field energy density and pressure, and dots and primes
denote, respectively, derivatives with respect to time and to the
field. By integrating the above equation, we also obtain the
following relation for the scalar field $\Phi$, i.e.,
\be
\label{phi11} \frac{\partial\Phi}{\partial a} =\sqrt{-\frac{m_{\rm
pl}^2}{8\pi
a}\frac{1}{\rho_{\Phi}}\frac{\partial\rho_{\Phi}}{\partial a}}\;,
\ee where $H\equiv\dot a/a$ stands for the Hubble parameter, and
the Friedmann equation $H^2={8\pi}\rho_{\Phi}/3m^2_{\rm pl}$ has
been explicitly used.

\begin{figure*}[t]
\centerline{\psfig{figure=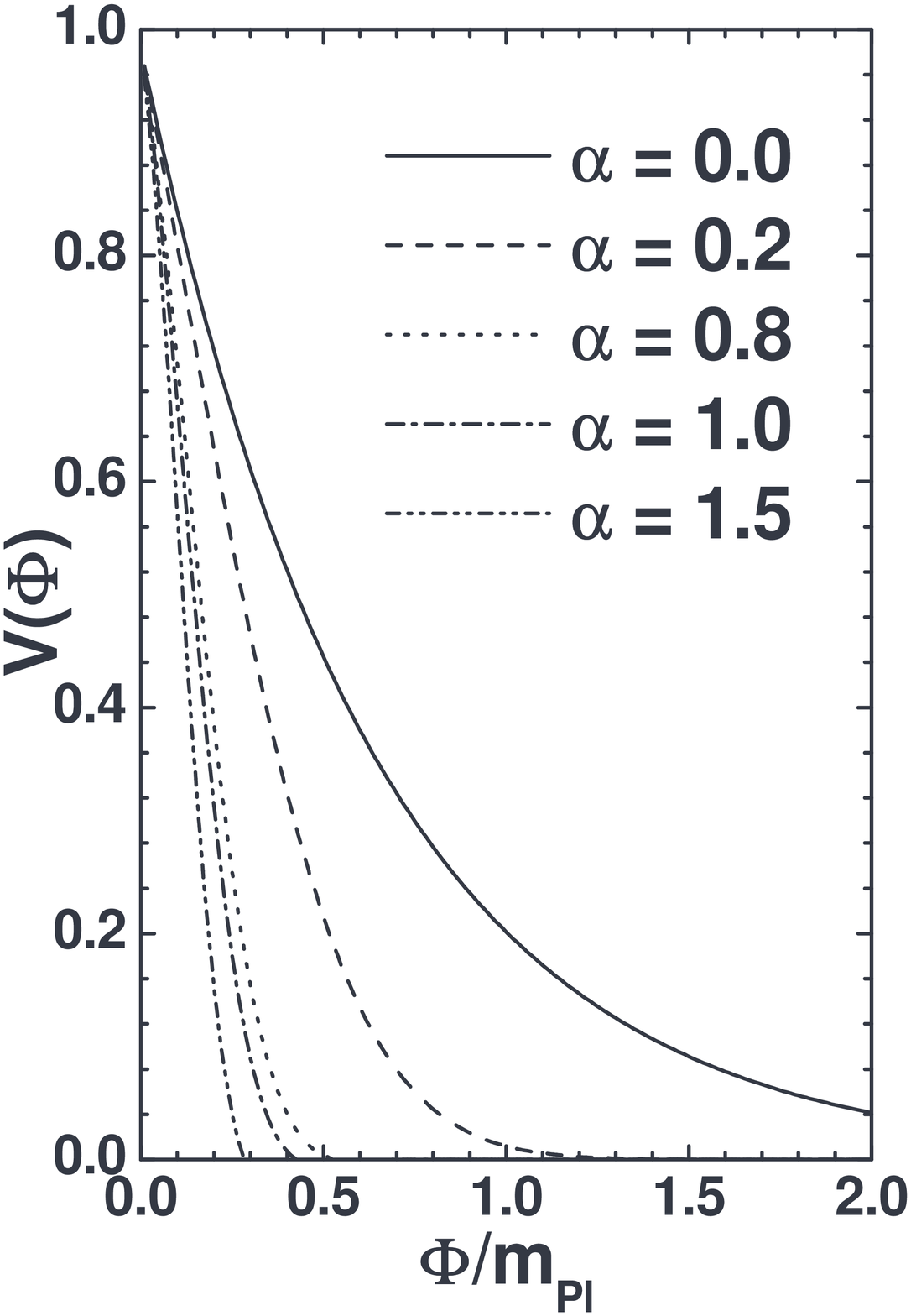,width=2.4truein,height=2.0truein}
\psfig{figure=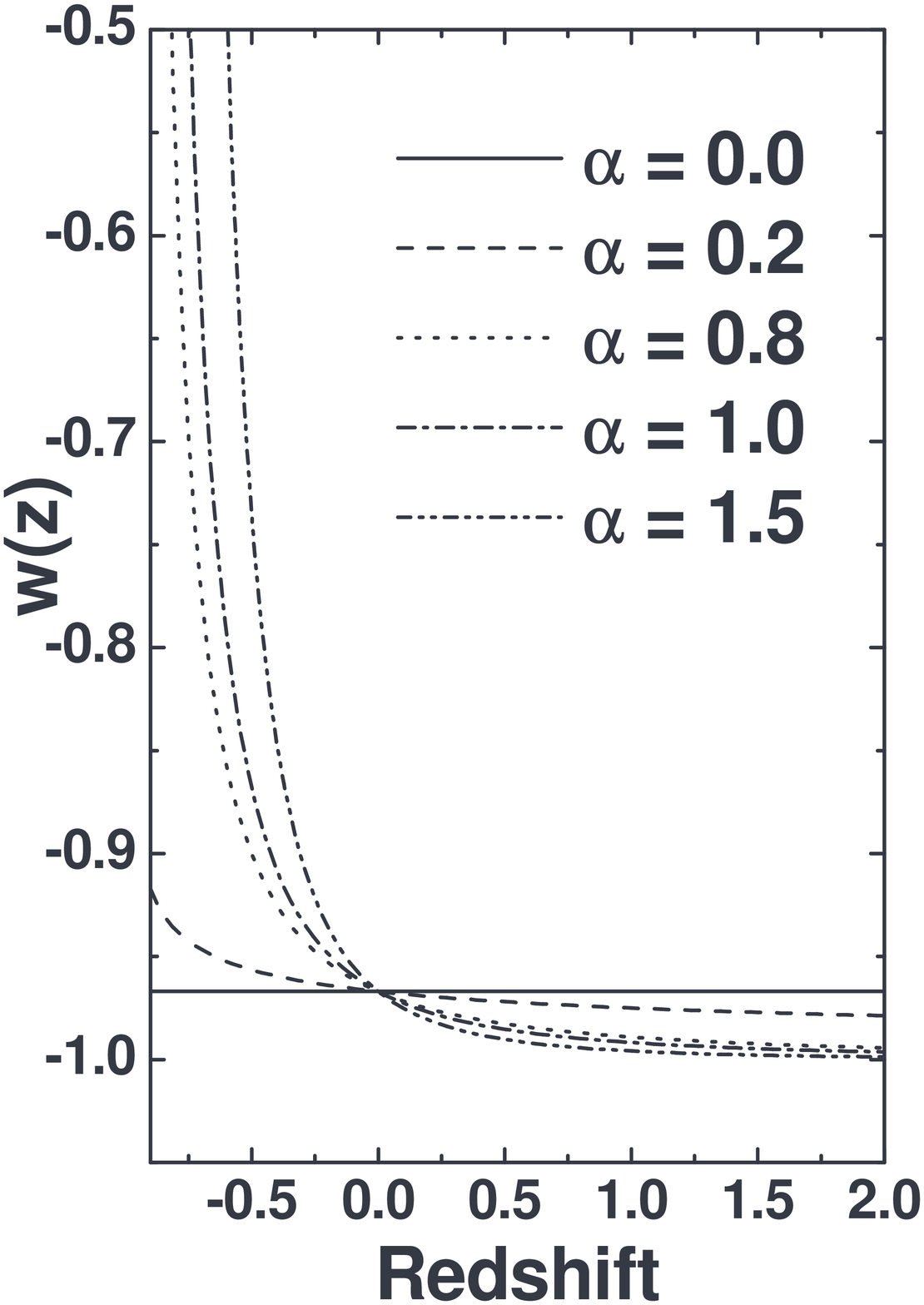,width=2.4truein,height=2.0truein}
\psfig{figure=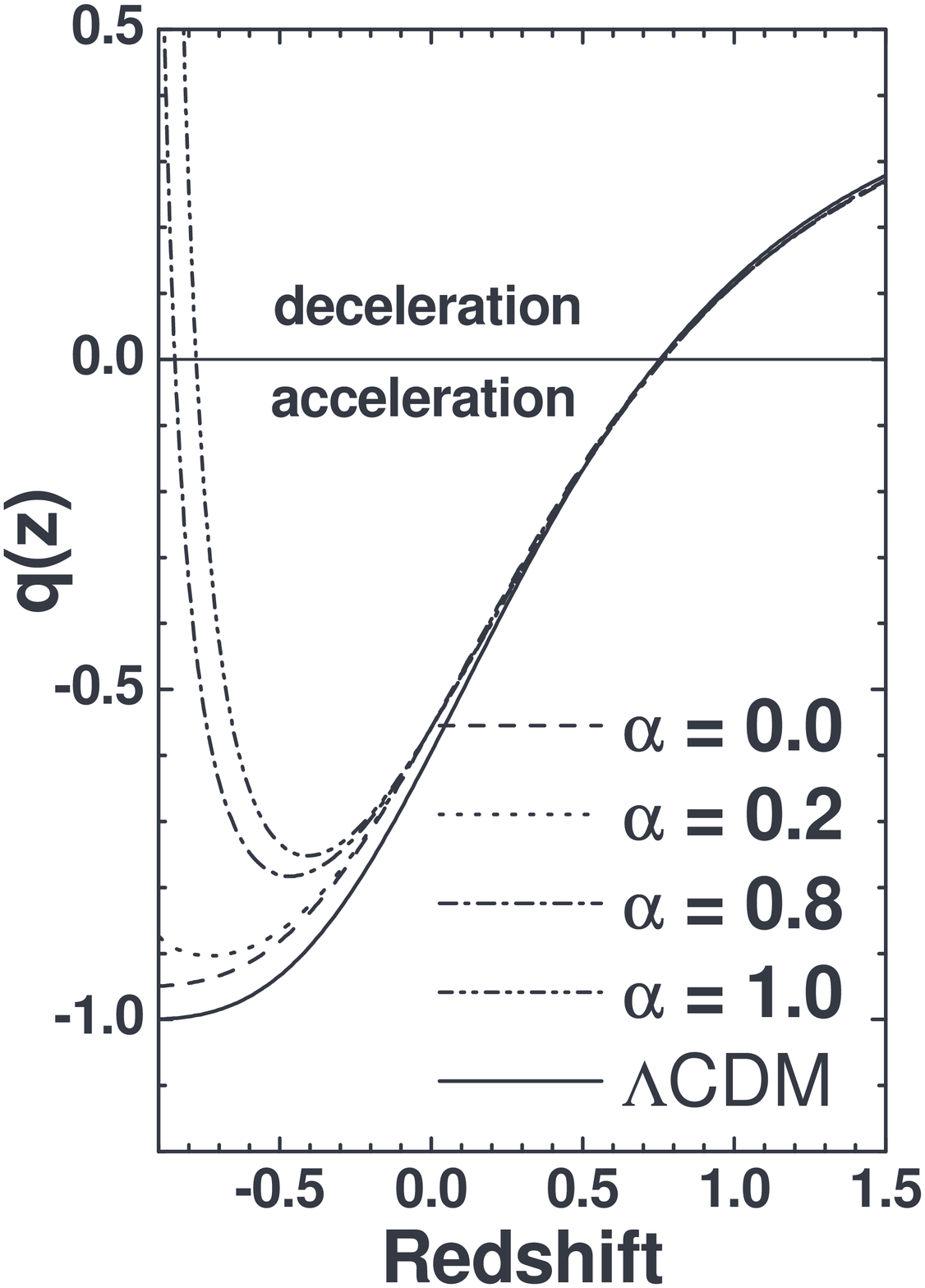,width=2.4truein,height=2.0truein} 
\hskip 0.1in} 
\caption{Some of the physical quantities discussed in the
text. {\bf{a)}} The potential $V(\Phi)$ as a function of the field
[Eq. (\ref{gpotential})] for some selected values of the parameter $\alpha$.
{\bf{b)}} The plane $w(z) - z$. Note that $w(z)$ reduces to a constant EoS $w
\simeq -0.96$ [$\lambda = {\cal{O}}(10^{-1})$] in the limit
$\alpha \rightarrow 0$ while $\forall$ $\alpha \neq 0$ it was $-1$
in the past and $\rightarrow +1$ in the future. {\bf{c)}} The
deceleration parameter as a function of the redshift for selected
values of $\alpha$ and $\Omega_{m,0} = 0.27$. For values of $\alpha \neq 0$ the cosmic
acceleration is a transient phenomenon. In particular, for $\alpha
= 1.0$ the transition redshifts happen at $z_{a/d} \simeq \pm
0.77$. The $\Lambda$CDM case (solid line) is also shown here for the sake of comparison.
}
\end{figure*}

In order to proceed further, let us adopt the following
\emph{ansatz} on the scale factor derivative of the energy density
\be
\label{ansatz}
\frac{1}{\rho_{\Phi}}\frac{\partial\rho_{\Phi}}{\partial
a}=-\frac{\lambda}{a^{1-2\alpha}}\;, 
\ee 
where $\alpha$ and $\lambda $ are positive parameters while the factor 2 was introduced for mathematical convenience. From a direct combination
of Eqs. (\ref{phi11}) and (\ref{ansatz}), the following expression for the scalar field is obtained 
\bea \label{phi1} \Phi(a) -
\Phi_0 = \frac{1}{\sqrt{\sigma}}\ln_{1-\alpha}(a)\;, 
\eea 
where $\Phi_0$ is the current value of the field $\Phi$ (from now on the subscript 0 denotes present-day
quantities), $\sigma = {8\pi/\lambda m_{\rm pl}^2}$ and the generalized function $\ln_{1 - \xi}$, defined as $\ln_{1 - \xi}(x)\equiv{(x^{\xi}-1)/\xi}$, reduces to the ordinary logarithmic function in the limit $\xi \rightarrow 0$ \cite{abramowitz}. To derive the potential $V(\Phi)$ for the above
scenario, we first note, from the definitions of $\rho_{\Phi}$ and
$p_{\Phi}$ [see Eqs. (\ref{density_phi}) and (\ref{eq_state})
below], that the potential $V(a)$ is given by 
\bea
\label{potential1} 
V(a)=\left[1-{\lambda\over6}a^{2\alpha}\right] \rho_{\Phi,0}\exp\left[-{\lambda \over
2}\ln_{1-\alpha}(a^2)\right]. 
\eea 
By inverting Eq. (\ref{phi1})
and inserting $a(\Phi)$ into the above expression, the potential
$V(\Phi)$ is readily obtained \footnote{Note that the inversion of
Eq. (\ref{phi1}) can be more directly obtained if one defines the
generalized exponential function as $\exp_{1-\xi}(x)\equiv[1 +
\xi{x}]^{1/\xi}$, which not only reduces to an ordinary exponential in the
limit $\xi \rightarrow 0$ but also is the inverse function of the
generalized logarithm ($\exp_{{1-\xi}}[{\ln_{1 -\xi}}(x)]=x$).
Thus, the scale factor in terms of the field can be written
as $a(\Phi)=\exp_{1-\alpha}[\sqrt{\sigma}(\Phi - \Phi_0)]$.}

\be
\label{gpotential} V(\Phi)= f(\alpha; \Phi)
\rho_{\Phi,0}\exp\left[-\lambda\sqrt{\sigma}\left(\Phi + {\alpha
\sqrt{\sigma} \over 2} \Phi^2 \right)  \right]\;, \ee
where $f(\alpha; \Phi) =
[1-{\lambda\over6}(1+\alpha\sqrt{\sigma}\Phi)^2]$. The important
aspect to be emphasized at this point is that in the limit $\alpha
\rightarrow 0$ Eqs. (\ref{phi1}) and (\ref{gpotential}) fully
reproduce the exponential potential studied by Ratra and Peebles
in Ref. \cite{RatraPeebles}, while $\forall$ $\alpha \neq 0$ the
scenario described above represents a generalized model which
admits a wider range of solutions. In order to exemplify this more
general behavior, Fig. (1a) shows the potential $V(\Phi)$ for some
selected values of the parameter $\alpha$ and the fixed value of
$\lambda = 10^{-1}$.

\vspace{0.1cm}

{\centerline{\emph{2. Scalar field + dark matter model.}}}

\vspace{0.1cm}

In what follows, it will be assumed that the cosmological fluid is
composed of non-relativistic matter (dark plus baryonic) and the
quintessence field $\Phi$. The Friedmann equation derived from the
action above now reads $H^2={8\pi/3m^2_{\rm pl}}\rho$, where $\rho =
\rho_{\Phi}+\rho_{m}$ is the total energy density.

A direct integration of Eq. (\ref{ansatz}) gives the scalar field
energy density as a function of the scale factor, i.e.,
\be
\label{density_phi}
\rho_{\Phi}(a)=\rho_{\Phi,0}\exp\left[-{\lambda \over
2}\ln_{1-\alpha}(a^2)\right]\;. \ee Note that in the limit $\alpha
\rightarrow 0$ the quintessence energy density (\ref{density_phi})
reduces to an usual power-law, $\rho_{\Phi}(a) \propto
a^{-{\lambda}}$ (as predicted by ordinary exponential potentials
\cite{RatraPeebles}), which clearly shows that this latter class
of $V(\Phi)$ provides only a particular solution out a set of
possible solutions that can be explored from a more general exponential law [see Eq. (\ref{gpotential})]. Note also that now the term $1/\rho_{\Phi}$ in the square root of Eq. (\ref{phi11})
must be replaced by $1/\rho$, so that by combining our
\emph{ansatz} (\ref{ansatz}) with the new Eq. (\ref{phi11}), we
find
\be
\label{phi2} \Phi - \Phi_0 = \frac{1}{\sqrt{\sigma}}\int_1^{a}{\frac{a'^{\alpha - 1}}{{\cal{F}}(a')}}da'\:, 
\ee
where ${\cal{F}}(a) = \sqrt{(1 + \Omega_m/\Omega_{\Phi})}$, with
$\Omega_m$ and $\Omega_{\Phi}$ representing the matter and
quintessence density parameters, respectively. As one may easily
check, the above expression for $\Phi(a)$ reduces to Eq.
(\ref{phi1}) for $\Omega_m = 0$. When combined numerically with
Eq. (\ref{potential1}), it also provides the potential $V(\Phi)$
for this realistic dark matter/energy scenario, which belongs to
the same class of potentials as given in Eq. (\ref{gpotential})
and shown in Figure (1a).

\emph{Equation of state.}---Without loss of generality to the
subsequent analyses, from now on we particularize our study to the
case $\lambda \simeq {\cal{O}}(10^{-1})$. Thus, by combining Eqs.
(\ref{eq_cons}) and (\ref{density_phi}), we also obtain the EoS
for this quintessence component, i.e., 
\bea 
\label{eq_state} 
w(a)= -1 + {1\over 30}a^{2\alpha}\;, 
\eea 
which is shown as a function
of the redshift parameter ($z = a^{-1} - 1$) in Fig. (1b) for some
selected values of the index $\alpha$. Differently from the
ordinary exponential cases studied in Ref. \cite{RatraPeebles},
the above EoS is a time-dependent quantity ($\forall$
${\cal{\alpha}} \neq 0$) and reduces to a constant EoS $w = - 1 +
{1 \over 30}$ [$\simeq -0.96$] only in the limit $\alpha
\rightarrow 0$, in agreement with our \emph{ansatz} (\ref{ansatz})
and the energy density derived in Eq. (\ref{density_phi}). Note
also that the EoS above (which must lie in the interval $-1 \leq
w(a) \leq 1$) is an increasingly function of time, being $\simeq
-1$ in the past, $\simeq -0.96$ today, and becoming more positive
in the future ($0$ at $a = 30^{1/2\alpha}$ and $1/3$ at $a =
40^{1/2\alpha}$). This amounts to saying that although the
Universe will be eternally dominated by the quintessence field
$\Phi$, it may not accelerate forever since the field will behave
more and more as an attractive matter field. Some physical
consequences of this unusual behavior are discussed as follows.

\emph{Transient Acceleration.}--- For a large interval of values
for  the parameter $\alpha$ the behavior of the EoS
(\ref{eq_state}) leads to a transient acceleration phase and, as a
consequence, may alleviate the dark energy/String theory conflict
discussed earlier. To study this phenomenon, let us first consider
the deceleration parameter, defined as $q =
-a\ddot{a}/\dot{a}^2$ and shown in Figure (1c) as a function of the
redshift parameter for some values of the index $\alpha$ and
$\Omega_{m,0} = 0.27$. As can be seen from this figure, $\forall$
$\alpha \neq 0$ the Universe was decelerated in the past, began to
accelerate at $z_* \lesssim 1$, is currently accelerated but will
eventually decelerate in the future. As  expected from Eq.
(\ref{eq_state}), this latter transition is becoming more and more
delayed as $\alpha \rightarrow 0$. In particular, at $a =
20^{1/2\alpha}$, $w(a)$ crosses the value $-1/3$, which roughly
means the beginning of the future decelerating phase. A
cosmological behavior like the one described above seems to be in
agreement with the requirements of String/M-theory (as discussed
in Refs. \cite{fischler}), in that the current accelerating phase
is a transitory phenomenon (Another interesting example of transient
acceleration is provided by the brane-world scenarios discussed in Ref. \cite{ss} (See also \cite{js}).). As one may also check, the
cosmological event horizon, i.e., the integral $\int{da/a^2H(a)}$
diverges for this transient scalar-field-dominated universe,
thereby allowing the construction of a conventional S-matrix
describing particle interactions within the String/M-theory
frameworks. A typical
example of an eternally accelerating universe, i.e., the
$\Lambda$CDM model, is also shown in Fig. (1c) for the sake of
comparison.

\begin{figure}[t]
\centerline{\psfig{figure=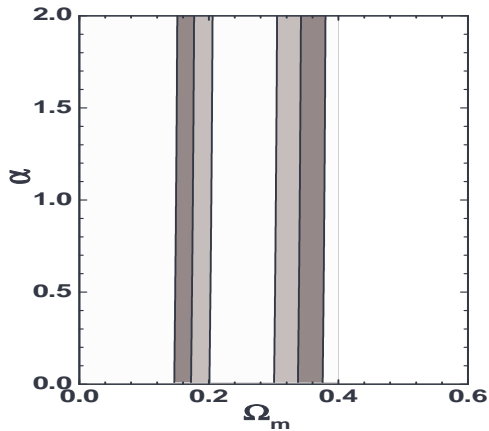,width=2.6truein,height=2.3truein}
\hskip 0.1in} \caption{68.3\%, 95.4\% and 99.73\% c.l. in the plane $\Omega_{m,0}$ - $\alpha$ arising from SNe Ia and CMB data. Note that while the matter density parameter is well restricted to the interval $\Omega_{m,0} = 0.25^{+0.06}_{-0.07}$ (at 95.4\% c.l.), the current observational data cannot place restrictive bounds on the parameter $\alpha$.}
\end{figure}

\emph{Observational Constraints.}---We study now some
observational bounds on the cosmological scenario proposed above.
We use to this end complementary data from the Supernova Legacy
Survey (SNLS) collaboration (corresponding to the first year
results of its planned five years survey) \cite{snls} and the
shift parameter from WMAP, CBI and ACBAR, defined as $R \equiv
\Omega_m^{1/2} \int_0^z{dz'/E(z')} = 1.716 \pm 0.062$, where $z =
1089$ is the redshift of recombination \cite{wmap}. The SNLS
sample used here includes 71 high-$z$ SNe Ia in the redshift range
$0.2 \lesssim z \lesssim 1$ and 44 low-$z$ SNe Ia compiled from
the literature but analyzed in the same manner as the high-$z$
sample. This data-set is arguably (due to multi-band, rolling
search technique and careful calibration) the best high-$z$ SNe Ia
compilation to date, as indicated by the very tight scatter around
the best fit in the Hubble diagram (we refer the reader to Ref. \cite{sne} for details on statistical
analyses involving SNe Ia and CMB data). Figure 2 shows the
68.3\%, 95.4\% and 99.73\% confidence limits (c.l.) in the
parametric space $\Omega_{m,0}$ - $\alpha$. Similarly to what
happens with most of the time-dependent EoS parametrizations (see,
e.g., \cite{sne}), the current observational bounds on the index
$\alpha$ are considerably weak since it appears only in the
exponential term of the energy density (\ref{density_phi}). We believe that the next generation of dark energy experiments dedicated to this issue (mainly those measuring the expansion history from high-$z$ SNe Ia, baryon oscillations, and weak gravitational lensing distortion by foreground galaxies -- see, e.g.,  \cite{jedi})
 will probe cosmology with sufficient accuracy to decide if values of $\alpha \neq 0$ are preferable from both theoretical and observational viewpoints 
(see also \cite{caldprl} and references therein for more on this issue). For the combination of current SNe Ia and CMB data, the best-fit model occurs for values of $\Omega_{m,0} = 0.25$ ($^{+0.06}_{-0.07}$ at 95.4\% c.l.) and $\alpha \simeq 1$  (with reduced $\chi_{\nu}^2 \simeq 1.01$), which corresponds to a 9.8$h^{-1}$-Gyr-old, accelerating universe with a deceleration parameter $q_0 =  -0.58$ and transition redshifts $z_a = 0.8$ (acceleration) and $z_d = -0.77$ (deceleration). A more detailed analysis of the cosmological model discussed here, as well as its conections with the inflationary scenario, will appear in a forthcoming communication.

\emph{Conclusions.}---We have constructed a model wherein the
quintessence field contributes as subdominant cosmological
constant at the earlier stages of the universe so that the
nucleossynthesis constraints  are naturally satisfied. However, although
subdominant for a long period (radiation and matter eras), the energy density of the field
$\Phi$ is increasing in the course of expansion, and, finally, for
a redshift of the order of a few, a quintessence dominated phase
begins. As we have discussed [see Eq. (\ref{eq_state})], a basic
difference with other quintessence models is that the accelerating
phase in the present scenario does not last forever. After some
eons, the equation of state parameter describing the field
component becomes more and more positive with the Universe,
inevitably, returning to an expanding decelerating stage. Finally,
we emphasize that the model makes definite predictions and is in
agreement with the observational tests analyzed here.

This work is supported by CNPq - Brazil. JSA is also supported by FAPERJ No. E-26/171.251/2004.


\begin{thebibliography}{99}

\bibitem{review} P.J.E. Peebles and B. Ratra Rev. Mod. Phys. {\bf{75}}, 559 (2003);
T. Padmanabhan, Phys. Rept. {\bf{380}}, 235 (2003); E. J. Copeland, M. Sami and S. Tsujikawa, hep-th/0603057.

\bibitem{weinberg} S. Weinberg, Rev. Mod. Phys. {\bf{61}}, 1 (1989). 


\bibitem{decaying}M. $\ddot{\rm{O}}$zer and M.O. Taha, Phys. Lett. B 171, 363 (1986);
K. Freese et al., Nucl. Phys. {\bf{B287}}, 797 (1987); J.C. Carvalho et al. J. A. S. Lima and I. Waga,  
Phys. Rev. {\bf{D46}}, 2404 (1992); J.A.S. Lima and M. Trodden, Phys. Rev. {\bf{D53}}, 4280 (1996).

\bibitem{quintessence}  R.R. Caldwell, R. Dave, P.J. Steinhardt, Phys. Rev. Lett. {\bf 80}, 1582 (1998); S.M. Carroll, Phys. Rev. Lett., {\bf{81}}, 3067 (1998); K.R.S. Balaji and R.H. Brandenberger, Phys. Rev. Lett. {\bf{94}}, 031301 (2005).

\bibitem{RatraPeebles}  B. Ratra and P.J.E. Peebles, Phys. Rev D{\bf 37},
3406 (1988).

\bibitem{fj} P.G. Ferreira and M. Joyce, Phys. Rev. D{\bf 58}, 023503(1998); A. Albrecht and C. Skordis, Phys. Rev. Lett. {\bf{84}}, 2076 (2000) 

\bibitem{wetterich} C. Wetterich, Astron. \& Astrophys. {\bf{301}}, 321 (1995).


\bibitem{power_law} P.J.E. Peebles and B. Ratra, Astrophys. J. Lett. {\bf 325}, L17 (1988); I. Zlatev, L-M. Wang , P.J. Steinhardt, Phys. Rev. Lett. {\bf{82}}, 896 (1999).

\bibitem{dodelson} S. Dodelson et al., M. Kaplinghat and E. Stewart,  Phys. Rev. Lett. {\bf{85}}, 5276 (2000)


\bibitem{fischler} W. Fischler, A. Kashani-Poor, R. McNees, and S. Paban, JHEP {\bf{3}}, 0107 (2001);
S. Hellerman, N. Kaloper and L. Susskind, JHEP {\bf{3}}, 0106,
(2001); J.M. Cline, JHEP 0108, 35 (2001);  E. Halyo, JHEP 0110, 025 (2001).

\bibitem{ms} M. S. Turner and A. G. Riess, Astrophys. J. {\bf{569}}, 18 (2002).

\bibitem{ph} R.R. Caldwell, Phys. Lett. {\bf{B545}}, 23 (2002). 

\bibitem{abramowitz} M. Abramowitz and I. Stegun, {\em Handbook of Mathematical Functions}, Dover, New York, (1965). See also J.A.S. Lima, R. Silva and A.R. Plastino, Phys. Rev. Lett., {\bf 86}, 2938 (2001).

\bibitem{ss} V. Sahni and Y. Shtanov, JCAP 0311, 014 (2003). 

\bibitem{js} J.S. Alcaniz and H. Stefancic, \texttt{astro-ph/0512622}.

\bibitem{snls} P. Astier {\it et al.}, Astron.\ Astrophys. {\bf 447}, 31 (2006).

\bibitem{wmap}  D.N. Spergel et al., \apj Suppl. {\bf{148}}, 175 (2003).


\bibitem{sne} T. Padmanabhan and T. R. Choudhury, MNRAS {\bf{344}}, 823 (2003); 
Y. Wang  and M. Tegmark, Phys. Rev. Lett., {\bf{92}}, 241302 (2004). M. A. Dantas et al., \texttt{astro-ph/0607060}.

\bibitem{jedi} A Crotts et al., \texttt{astro-ph/0507043}. (see also http://jedi.nhn.ou.edu/).

\bibitem{caldprl} R.R. Caldwell and E. V. Linder, Phys. Rev. Lett. {\bf{95}}, 141301 (2005).

\end{thebibliography}
\end{document}